\documentclass[twocolumn,nofootinbib]{revtex4}

\usepackage{graphicx}
\usepackage{color}
\usepackage{url}
\usepackage[utf8]{inputenc}
\usepackage{natbib}

\usepackage[
       colorlinks=true,
       urlcolor=blue,       
       filecolor=green,     
       linkcolor=red,       
        bookmarksopen=true]{hyperref}

\usepackage{amsmath,amssymb}

\def\lsim{\raise0.3ex\hbox{$\;<$\kern-0.75em\raise-1.1ex\hbox{$\sim\;$}}}
\def\gsim{\raise0.3ex\hbox{$\;>$\kern-0.75em\raise-1.1ex\hbox{$\sim\;$}}}

\def\be{\begin{equation}}
\def\ee{\end{equation}}
\def\nn{\nonumber}

\def\theta{\vartheta}

\def\R{{\cal R}}

\def\ncre{n_{\rm cr,e}}

\def\d{{\rm d}}

\renewcommand{\vec}[1]{\boldsymbol{#1}}

\newcommand\vev[1]{{\left\langle{#1}\right\rangle}}

\definecolor{purple}{rgb}{0.5,0,0.5}

\definecolor{cadmiumgreen}{rgb}{0.0, 0.42, 0.24}
\definecolor{evilblue}{rgb}{0.05,0.05,0.35}

\newcommand{\aap}{{Astron.\ Astrophys. }}

\begin{document}

\title{Polarized synchrotron data and the structure of the Galactic magnetic field}

\author{C.~T.~Becker}
\author{M.~Kachelrie\ss}

\affiliation{Institutt for fysikk, NTNU, Trondheim, Norway}

\date{\today}

\begin{abstract}
The polarized synchrotron data of the northern C-BASS survey show a
surprisingly low linear polarization fraction, $\vev{\Pi}\simeq 3\%$, while
the magnetic field polarization features coherent structures over rather
large angular scales. The low polarization degree points to a strong
dominance of the turbulent magnetic field---in agreement with the observation
that the total synchrotron intensity is much larger than expected
from the regular component in current models for the Galactic magnetic
field (GMF). In contrast, studies of cosmic ray propagation employing these
GMF models suggest that cosmic rays propagate anisotropically, what in turn
requires weak turbulent fields.
As a solution to this contradicting requirements, we suggest that the GMF
consists of three components with different levels of turbulence: a disk
dominated by the turbulent field, a halo  dominated by the regular field,
plus an extended turbulent halo field.
\end{abstract}


\date{August 3, 2024}

\maketitle

\section{Introduction}

Galaxies are permeated by magnetic fields with strengths of the order of
$\mu$G. These magnetic fields have not only an major impact on the evolution
of galaxies but are also responsible for the deflection and the confinement
of cosmic rays (CR)~\cite{Beck:1995zs,Boulanger:2018zrk}.
In turn, the synchrotron emission of CR electrons is
a key signature both of CRs themselves and of magnetic fields. In general,
magnetic fields  can be split into a regular,
average component $\vec{B}$ and a chaotic, turbulent component $\vec b$
with $\vev{\vec b}=0$. While the average $\vev{\cdot}$ is formally an
ensemble average, observationally it has to be replaced by
an average over, e.g., different regions which are statistically
independent and subject to similar physical conditions.

Faraday Rotation Measures (RMs) of Galactic and extragalactic polarized
radio sources provide important information on the Galactic magnetic field
(GMF). They are proportional to the line-of-sight integral of the product
of the thermal electron density $n_e$ of the interstellar medium and the
magnetic field parallel to the line-of-sight,
$
\text{RM} \varpropto \int_0^L n_e(s)\, [B_\parallel(s)+b_\parallel(s)] \,
\text{d}s,
$
where $L$ is the distance to the source. 
Another important source of information on the GMF is the synchrotron
radiation emitted by relativistic CR electrons. Sky maps of the
Stokes parameters $Q$ and $U$ of the polarized synchrotron emission from
the Galaxy were extracted from the data of the WMAP and Planck satellites.
These parameters encode the
orientation of the transverse component of the magnetic
field. The strength of the polarized emission,
$P = \sqrt{{Q}^2 + {U}^2}$ is again given by a line-of-sight
integral, approximately given by
$
P \varpropto \int_0^L \ncre(s)\, [B_\perp^2(s)+b_\perp^2(s)] \, \text{d}s,
$
where $n_\text{cre}$ denotes the density of cosmic-ray electrons.
The total intensity, $I$, follows from a similar relation.

Thus, RM and radio sky maps provide complementary data on
$B_\parallel$ and $B_\perp$ and, having enough tracers along the
line-of-sight,
one may attempt to reconstruct the three-dimensional structure of the
GMF. Assuming, as it is often done, that the
contribution of the turbulent field $\vec b$ cancels completely
in the RM and the polarized intensity $P$ allows one to disentangle
also the regular and turbulent components of the field.
However, the morphology of current models of the GMF like those of
Refs.~\cite{Pshirkov:2011um,Jansson:2012pc,Jaffe:2013yi,Unger:2023lob,Korochkin:2024yit} differ substantially,
although they reproduce the Faraday RM data and synchrotron emission maps 
to which they were fitted.
This implies in particular that it is at present not possible to derive a
unique best-fit model for the GMF. In order to illustrate at least partially
the currently allowed parameter space of these models, the authors of
Ref.~\cite{Unger:2023lob} presented therefore recently not a single
but a suite of eight GMF models.

Another sign for the inconsistency of current GMF models is that fits
to either RMs or synchrotron data yield conflicting results:
The overall magnetic field strengths derived from RMs are a factor few
smaller than from synchrotron emission.  Explanations
for this discrepancy include the existence of anisotropic random
fields contributing to $P$ but not to RM, or an anti-correlation of
$n_e$ and $B$ leading to an underestimation of $B$ from
RMs~\cite{2003A&A...411...99B}.
An additional source of uncertainties in the GMF models are
deficiencies in the models for the three-dimensional densities of
thermal and CR electrons. In particular, the latter are not
modeled self-consistently together with the GMF but are derived
using an independent and over-simplified diffusion frame-work. 

In this work, we do not aim to improve relative to existing GMF
models. Instead, we show how three seemingly contradicting
observations---the recent polarization data from the C-Band All-Sky Survey
(C-BASS)~\cite{Paddy,Singal:2022jaf}, constraints from CR
escape~\cite{Giacinti:2014xya,Giacinti:2015hva,Giacinti:2017dgt}, and
the slow diffusion found close to some TeV pulsar-wind nebula (PWN) like
Geminga~\cite{Abeysekara:2017old}---can be explained and matched to
different elements of a schematic GMF model. After reviewing these
observations, we present our schematic model and derive the expected
polarization degree towards the Galactic north pole and the latitude
profile of the synchrotron intensity. We obtain an acceptable agreement
with the data for a model consisting of three components with different
levels of turbulence: a disk dominated by a turbulent field with small
coherence length, a halo  dominated by the regular field,
plus an extended turbulent halo field.

\section{C-BASS polarization data}

The C-BASS collaboration aims to obtain an all-sky map of the
Galactic synchrotron  emission at 5\,GHz in both total intensity and
polarization. Preliminary results of the northern C-BASS instrument were
presented in Refs.~\cite{Paddy,Singal:2022jaf}.
The minimal distortions caused by Faraday rotation at this
frequency and the higher signal-to-noise ratio compared to the WMAP and
Planck surveys made 
for the first time an accurate separation of the synchrotron from other
emission processes possible. In the northern high latitudes, $b>30^\circ$,
and excluding Loop\,I and III, an average linear polarization degree of
$\vev{\Pi}\equiv \vev{P/I}\simeq 3.3\%$ was reported, with only few pixels
exceeding 10\%.
This low values should be contrasted with the expectation of
$\Pi=(1+p)/(7/3+p)= 75\%$ for a regular, uniform magnetic field
and a power-law energy  distribution $\d N/\d E\propto E^{-p}$ of CR electron
energies with $p=3$.

In addition, results for the two-point angular correlation function
$D(\theta)=\vev{\chi(\vec n)-\chi(\vec n')}$ of the relative orientation $\chi$
of the local polarization basis were presented in
Refs.~\cite{Paddy,Singal:2022jaf}. The shape of
the two-point correlation function $D(\theta)$, where $\theta$ is the
angle between the unit vectors $\vec n$ and $\vec n'$,
indicates that the typical angular size of field structures is around
$\theta_0\simeq 15^\circ$.
Using a simple cell model for a purely turbulent field, where one identifies
the cell size $L$ with the coherence length $L_c$ of the turbulent field,
one expects $\sin(\theta_0/2)\simeq 1/N$ for a a line-of-sight with length $NL$.
Assuming then a random walk for the orientation $\chi$, the scale
$\theta_0\simeq 15^\circ$ implying ${\cal O}(N)\sim 10$ cells leads to
$\Pi\simeq 0.75/\sqrt{10}\simeq 24\%$~\cite{Paddy}. Thus there is a
contradiction between
the observed coherence of the magnetic field polarization over rather
large angular scales and the low polarization degree in the C-BASS data.
Adding a regular field would strengthen the discrepancy.

We can quantify this disagreement calculating the linear polarization
degree for the suite of GMF models  recently presented in
Ref.~\cite{Unger:2023lob}, adding a turbulent random field normalized
by $b(\vec x)=\eta B(\vec x)$.
We choose the coherence length as $L_c=20$\,pc and $p=2.8$ as slope
of the electron energy distribution.
In Table~\ref{tab_pol}, we report the expectation value of $\vev{\Pi}$  towards
the Galactic north pole. We see that the variation in $\vev{\Pi}$  between
different models for the regular field is very small. Increasing $\eta$,
the average polarization degree decreases until it flattens out at
$\vev{\Pi}\simeq 0.12$ for $\eta\gg 1$. Even lower  polarization levels
would require smaller
values of the coherence length. Thus the measured  polarization degree
requires a strong dominance of the turbulent field and small coherence
lengths, in contradiction to the large angular size of
the observed polarization features.

\begin{table}
\begin{tabular}{ccccccccc}
        \hline
        $\eta$ & \texttt{base} & \texttt{neCL} & \texttt{expX} & \texttt{spur} & \texttt{cre10} & \texttt{synCG} & \texttt{twistX} & \texttt{nebCor} \\
        \hline
        1   & 0.36 & 0.37 & 0.36 & 0.39 & 0.33 & 0.39 & 0.36 & 0.35 \\
        1.5 & 0.22 & 0.23 & 0.22 & 0.24 & 0.2 & 0.24 & 0.22 & 0.21 \\
        10 & 0.12 & 0.12 & 0.12 & 0.12 & 0.12 & 0.12 & 0.12 & 0.12 \\
        \hline
\end{tabular}
\caption{The average linear polarization degree $\vev{\Pi}$ for the eight GMF
  models from Ref.~\cite{Unger:2023lob} 
  plus a turbulent field with strength  $b=\eta B$. \label{tab_pol}}
\end{table}

\section{Cosmic ray escape}

\subsection{Isotropic diffusion}

The large majority of studies on CR propagation in the Milky Way employs a
phenomenological approach, using an isotropic diffusion coefficient $D(\R)$
uniform in a cylinder enclosing the Galactic disk. Its  dependence on
rigidity $\R=cp/(ZeB)$ is typically parameterised by a (broken) power law,
$D(\R)= D_0 (\R/\R_0)^{-\delta}$.
Then the parameters $D_0$ and $\delta$ are determined from fits to
secondary-to-primary ratios as, e.g., boron-to-carbon and the
abundance of radioactive secondaries as $^{10}$Be/$^9$Be.
Typical values found from a combined fit to many observables
using GALPROP~\cite{Johannesson:2016rlh,Porter:2021tlr} are in the range
$(5-8)\times 10^{28}$cm$^2$/s at the reference rigidity $\R_0=4$\,GV with
the high-energy slope $\delta=0.34-0.36$. Note that this slope is very close to
the expectation $\delta=1/3$ for a Kolmogorov power spectrum
of the turbulent magnetic field.
Similar values for $D_0$ and $\delta$ are found using other
CR diffusion codes like, e.g.,
DRAGON~\cite{Evoli:2008dv,delaTorreLuque:2022vhm}.

In the picture of an uniform diffusion coefficient, one can model the
propagation of CRs as a random walk with an energy dependent effective step
size $L_0$. For a pure isotropic random field, one expects  as functional
dependence of the diffusion coefficient
\be  \label{diff}
  D_{\rm iso} = \frac{cL_0}{3} 
 \left[ (R_{\rm L}/L_0)^{2-\gamma} + (R_{\rm L}/L_0)^2 \right] ,
\ee
where  $R_{\rm L}=\R/B=cp/(ZeB)$ is the Larmor radius of the CR and 
the condition $R_{\rm L}(E_{\rm tr})=L_0$ determines the transition from 
small-angle scattering with $D(E)\propto E^2$ to
large-angle scattering with $D(E)\propto E^{2-\gamma}$ and
$\gamma=5/3$ for a Kolmogorov power spectrum.

The numerical value of $L_0$ should scale with the correlation length 
as $L_0\propto L_{\rm c}$, but the proportionality factor has to be
determined numerically. In Refs.~\cite{Parizot:2004wh,Giacinti:2017dgt},
it was found that $L_0 \simeq L_{\rm c}/(2\pi)$ provides a good 
fit to their numerical results for isotropic Kolmogorov turbulence.
For values of the field strength and
correlation length expected for the GMF, the predicted diffusion
coefficient from Eq.~(\ref{diff}) is a factor of order 100 smaller
than those extracted using,
e.g., Galprop~\cite{Johannesson:2016rlh,Porter:2021tlr} or
DRAGON~\cite{Evoli:2008dv,delaTorreLuque:2022vhm}.
A similar discrepancy was found earlier performing analytical estimates,
but was considered to be within the error range expected for the
approximations made~\cite{Strong:2007nh}.

However, a factor 100 difference in the diffusion coefficient leads to
dramatic consequences for the GMF parameters: The weak field dependence of
$D(E)\propto B^{-1/3}$ requires a reduction of $B$ by a
factor $1/100^3=10^{-6}$ for Kolmogorov turbulence keeping $L_{\rm c}$ constant.
Keeping instead  $B\sim {\rm few}\times\mu$G, the correlation length
should be comparable to the size of the Galactic halo. 
Therefore, it was concluded in Ref.~\cite{Giacinti:2017dgt} that
CR escape is inconsistent with isotropic diffusion. This implies in turn
that the regular field should dominate in the regions relevant for the
CR escape from the Galactic disk.

\subsection{Anisotroic propagation in GMF models}

For a given GMF model, the propagation of CRs in the Milky Way can be
studied also directly integrating the Lorentz equation. Calculating then
observables like the
grammage crossed by CRs or their escape time allows one to cross-check
the viability of this GMF model. Since this approach is at low energies
computationally expensive, it has been restricted to either relatively high
energies~\cite{DeMarco:2007eh} or
young sources~\cite{Kachelriess:2015oua}.
In Refs.~\cite{Giacinti:2014xya,Giacinti:2015hva}, 
the trajectory approach was applied for the first time to calculate the
diffuse flux of CR primaries below the knee, using as  lowest rigidity
$\R=10^{14}$\,V. The authors calculated the grammage $X$ crossed by CRs
before escape in the Pshirkov et al.~\cite{Pshirkov:2011um} and
Jansson-Farrar (JF) models~\cite{Jansson:2012pc,Jansson:2012rt}
for the regular GMF together with
a turbulent component with a Kolmogorov power-spectrum.
They confirmed at the lowest rigidities the dependence
$X\propto \R^{-1/3}$ expected in the large-angle scattering regime for
Kolmogorov turbulence. This allowed them to connect their results to the
observations at lower energies, especially to those from the AMS-02
collaborations~\cite{Aguilar:2016vqr}. In order to obtain agreement with
these measurements, they had to reduce the strength of the turbulent field.
As a result of this reduction, the low anisotropy level lead
to strongly anisotropic propagation of CRs.

Anisotropic diffusion by itself, however, is not enough to facilitate the CR
escape from the Galaxy: If the GMF model does not contain a poloidal
component, only the CR transport along the plane towards either the Galactic
bulge or its outer radial edge would be increased---what is too inefficient
to reduce the
grammage significantly~\cite{DeMarco:2007eh}. This restriction is however
not severe, since any realistic GMF model should contain a poloidal component,
induced e.g.\ by Galactic winds and outflows. In the case of the JF model,
its ``X-field'' leads to an additional tilt of the total magnetic field
with respect to the Galactic plane.

\begin{figure}
\includegraphics[width=0.95\columnwidth]{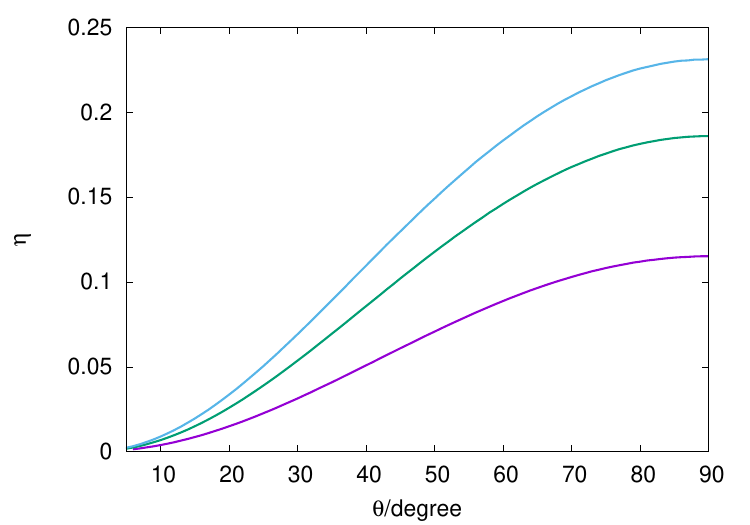}
\caption{The anisotropy level $\eta$ as function of the tilt angle $\theta$
  required for $D_{zz}=D_{\rm iso, obs}$, for $\{L_{\max}/{\rm pc},B/\mu {\rm G}\}$
  equal to $\{200,3\}$ (top), $\{100,3\}$ (middle) and $\{100,4\}$ (down).
   \label{fig:theta}}
\end{figure}

We can estimate the level of anisotropy required to obtain consistency
with the isotropic diffusion coefficient obtained fitting secondary
ratios considering a simple toy model~\cite{Giacinti:2017dgt}.
If the regular magnetic field in 
the halo has on average the tilt angle $\theta$ with the Galactic
plane, the component of the diffusion tensor relevant for CR escape is given
by 
\be \label{Dzz}
D_{zz} = D_\perp\cos(2\theta) + D_\|\sin(2\theta) . 
\ee
Requiring then $D_{zz}=D_{\rm iso}=(5-8)\times 10^{28}$cm$^2$/s as obtained
with GALPROP~\cite{Porter:2021tlr} at
$\R_0=4$\,GV fixes $\eta$ as function of the tilt angle $\theta$.
In Fig.~\ref{fig:theta}, we show  the resulting $\eta(\theta)$  relation for
three different sets of values of the  parameters
$L_{\max}$ and $B$. Avoiding too small values of $\eta$ clearly
favors relatively large values of the coherence length. Still, the
turbulence level $\eta$ should be around or below 10\% for a tilt
angle $\theta\simeq 30^\circ$. Note that this level corresponds approximately
to the one found propagating individual CRs in the JF model in
Refs.~\cite{Giacinti:2014xya,Giacinti:2015hva}. Moreover,  the tilt of the
total magnetic field with the Galactic plane induced by the X-field
varies in the JF model between $\theta\simeq 10^\circ$ at Earth, increasing to
$\theta\simeq 50^\circ$ further away from the Galactic plane.
Thus an average tilt angle of $\theta\simeq 20^\circ$--$30^\circ$ is a
reasonable approximation to this model. 

\subsection{Isotropic diffusion close to PWNe}

The diffusion coefficient deduced by the HAWC~\cite{Abeysekara:2017old}
collaboration for  Geminga and PSR~B0656+14 corresponds to
$D=(2.1\pm 0.6)\times 10^{26}$cm$^2$/s  at 10\,GV,  scaling to low energies
using Kolmogorov turbulence. 
We can compare this value with the one predicted by the fitting
formula~(\ref{diff}) valid in the case of purely isotropic turbulence.
Inserting numerical values appropriate for the HAWC measurement, it
follows
\be  \label{iso}
D_{\rm iso} \simeq 2\times 10^{26}\frac{\rm cm^2}{\rm s}
\left( \frac{\R}{10{\rm GV}}\right)^{1/3}
\left( \frac{L_c}{2{\rm pc}}\right)^{2/3} 
\left( \frac{B}{4\mu{\rm G}}\right)^{-1/3} .
\ee
Thus the numerical value of the diffusion coefficient
deduced from HAWC observations agrees with values expected from isotropic
diffusion, for field strengths  typical for the Galactic plane but relatively
small coherence lengths.

This finding is in line with the results of
Ref.~\cite{Lopez-Coto:2017pbk}. These authors followed the trajectories
of individual electrons in purely isotropic turbulence and compared
the resulting gamma-ray emission to HAWC data, finding as best-fit
value $b\simeq 3\mu$G and small coherence lengths, $L_c\lsim 5$\,pc.
In principle, larger coherence lengths can be accommodated in Eq.~(\ref{iso})
by increasing appropriately the field strength. However, as it was stressed
first in Refs.~\cite{Giacinti:2012ar,Giacinti:2013uya}, CR propagation
proceeds on scales smaller than few$\,\times L_c$ even in isotropic turbulence
anisotropically, leading to a characteristic anisotropic source morphology.
This argument was used in Ref.~\cite{Lopez-Coto:2017pbk} to exclude larger
coherence lengths.

Next we ask if one can avoid isotropic diffusion close to PWNe.
Allowing for the presence of a regular magnetic field component close
to PWNe leads to two consequences: On one hand, the  perpendicular diffusion
coefficient decreases, $D_\perp \leq  D_{\rm iso}$,  opening in principle the
possibility to reach the values indicated by Eq.~(\ref{iso}). On the other
hand, the parallel diffusion coefficient increases even faster, resulting
in anisotropic diffusion and thus in general asymmetric source
morphologies.
Thus the problem of this scenario are again the resulting asymmetric source
morphologies.
Already for $\eta=0.5$, the ratio $D_\|/D_\perp \simeq 100$ at
100\,TeV~\cite{Giacinti:2017dgt}
would imply a strongly anisotropic source morphology --
with the source extension along the regular field direction much larger
than deduced from HAWC observations. For a single source, such an asymmetry
can be avoided if our line-of-sight to the source is by chance aligned
with the regular magnetic field at the source. In the case of Geminga, one
expects however a rather large angle between the line-of-sight and the regular
magnetic field for any GMF model following mainly a spiral-like structure
in the disk.
Moreover,  a search for asymmetries in the TeV halos around PWNe was
conducted using Fermi-LAT data in Ref.~\cite{DelaTorreLuque:2023ssz}.  
No evidence for asymmetries was found. 
We thus conclude that observations favor isotropic diffusion
of CRs close to PWNe, with values of the isotropic diffusion coefficient
in line with expectation from simulations of test particle trajectories
in such fields.

\section{Structure of the Galactic magnetic field}

\subsection{Three-component model}

Let us now try to match the seemingly contradicting observations to different
elements of a schematic GMF model. For this model, we suggest three
main components, a disk, a halo and an extended halo which we will also call
corona field:
\begin{itemize}
\item
  In order to ensure sufficient fast CR escape, we have to assume that the
  halo field is dominated by the regular field and that the sub-dominant
  turbulent component has a large coherence
  length $L_c$, such that $\sin(2\theta) D_\|\simeq D_0$ holds. 
\item
  In the disk, we assume that, at least locally, the magnetic field is
  dominated by the turbulent component and has a small $L_c$,
  such that $D\simeq D_{\rm iso}\simeq D_0/100$ is valid.
  In this case, the number $N$ of cells with coherence length $L_c$
  along a line-of-sight towards the Galactic North pole is  $N={\cal O}(10)$
  in the disk and, thus, there will be only a partial cancellation of the
  contribution of the turbulent disk field to the polarization.
  \\
  The dominance of the turbulent field could be either only a local anomaly,
  meaning that in the Local Bubble (and other super bubbles as well as close
  to some PWNe) diffusion is slow
  but otherwise not. Or, it could hold globally or at least in a large
  fraction\footnote{Note that the authors of Ref.~\cite{Jacobs:2023zch}
      have argued that the filling factor of slow-diffusion regions in the
      Galactic disk is around 2/3.} of the disk. 
  Since we test only line-of-sights starting at
  the Earth towards the Galactic North pole, we cannot distinguish between
  these two possibilities.
\item
  Finally, we add an extended halo field: This field is also dominated by
  the turbulent component, but has a larger $L_c$, and may extend up to or
  beyond 200\,kpc. This component will not contribute to the polarization,
  while adding emissivity to the total synchrotron intensity.
  In addition to the C-BASS data, the presence of such a halo is
    motivated both by observations and simulations of the large-scall
    structure which show that
    magnetic fields and CR electrons typically fill entire galaxy
    clusters~\cite{Botteon:2022umz}.
\end{itemize}

In the following, we will not attempt to construct a detailed GMF model
consisting of these three components. Instead, we will use these
three components as a toy model to test if we can reproduce qualitatively
the synchrotron and polarization data.

\subsection{CR electron distribution}

In addition to the magnetic field model, we require a model for the
CR electron distribution  $\ncre(E,\vec x)$. For the energy dependence and
the normalization of the CR electron distribution in the Galactic disk, we use
the local interstellar spectrum derived from
the Voyager and AMS-02 data in Ref.~\cite{Bisschoff:2019lne}.
In order to have simple formula for the
synchrotron emissivity, we approximate the energy dependence by a
power law, $\d N/\d E\sim E^{-p}$. Moreover, we assume factorization,
$\ncre(E,\vec x)=n(\vec x) \d N/\d E$, and use an one-dimensional
spatial picture. Thus we ignore a radial dependence of $n(\vec x)$ and
solve only for the $z$ dependence. For an extended halo field with strength
$B\simeq 0.5\mu$G, the characteristic electron energy to produce 5\,GHz
synchrotron radiation is $E_{\rm cr}\simeq 12$\,GeV. At these energies,
advection, energy losses due to synchrotron radiation and inverse Compton
scattering, and energy gain due to re-acceleration play a role. While
a study including all these effects is beyond the scope of this work, we
note that the energy losses $E\d E/\d t\simeq-2\times 10^{-14}$\,GeV$^2$/s
are balanced by energy gains $E\d E/\d t=c^2 D_{\rm pp}\simeq (2Ev_A/3)^2/D(E)$
for Alfv\'en velocities around 10\,km/s. Determining therefore for simplicity
$n(z)$ from a stationary 1d diffusion equation,
the solution is
\be  
n(z)=n_0 - j_0\int_0^z  \frac{\d z'}{D(z')} 
\ee
for a prescribed profile $D(z)$ of the diffusion coefficient. 
Here, $j_0$ and  $n_0$ denote the current and the CR electron number density
in the Galactic disk, respectively.
Choosing an exponential profile for the diffusion coefficient,
$D(z)= D_0 \exp(z/z_0)$, the solution becomes
\be \label{eq:n}
 n(z)= n_0 +\frac{j_0 z_0}{D_0}  \left[ \exp(-z/z_0) -1 \right] .
\ee
This solution has two qualitatively different cases: For
$j_0 z_0/D_0n_0>1$, the point $n(z_\ast)=0$ defines a free-escape boundary
at $z_\ast$, 
while for $j_0 z_0/D_0n_0<1$ the density  $n(z)$  stays non-zero for
$z\to\infty$.

In the subsequent fits, we will fix the $z$ dependence of the CR
electron density to the one from of Eq.~(\ref{eq:n}), setting 
$j_0 z_0/(D_0 n_0) \simeq 2.0$ together with a minimum value of
$n_\text{min} = n(z_c)/n_0 = 0.1$.
For the magnetic field in the disk, we assume that it is constant with
strength $B_0$  and decreases outwards as  $B(z)\propto \exp[-(z-z_0)/z_B]$
for $z > z_\text{disk}$ until it reaches the minimal value $B_\text{min}$.
Inspired by the relation  $B \propto D^{-1/3}$ for Kolmogorov turbulence,
we connect the scale heights of the magnetic field to the one of the electron
density by $z_{B} = z_{0}/3$.

\section{Results}

\subsection{Model parameters and fit procedure}

Even our simple toy model contains the 13~parameters shown in
Table~\ref{tab:parameters_dhc} that have to be determined in principle
from observations.
Our aim is however not to find ``the best-fit'' model, but only to
check if a consistent explanation both of the low polarization degree
and the synchrotron data is possible. We fix therefore a number of parameters.
In particular, we choose for the vertical extension, the coherence lengths,
and the vertical scale heights of the electron distribution
and the magnetic field the values shown in Table~\ref{tab:DHC_setup}.
Moreover, we fix the normalization of the electron number density,
varying only the slope $p$.
For the regular halo field, we neglect the sub-dominant turbulent field
and use only a homogeneous field. To add  small variations, we divide the halo
in 10~patches of size $H_\text{halo}/10$. In each patch, we choose the field
direction picking a vector within a cone with opening angle $\theta=15^\circ$
around the initial direction of the homogeneous field.

\begin{table}[!htb]
    \centering
    \begin{tabular}{ll}
        \hline
        Parameter & Symbol \\
        \hline
        Correlation length of corona & $L_{c,\text{corona}}$ \\
        Correlation length of disk & $L_{c,\text{disk}}$ \\
        Correlation length of halo & $L_{c,\text{halo}}$ \\
        Disk field strength & $B_0$ \\
        Electron number density normalization constant & $n_0$ \\
        Electron number density scale height & $z_{0}$ \\
        Magnetic field strength scale height & $z_{B}$ \\
        Minimum electron number density fraction & $n_\text{min}$ \\
        Minimum magnetic field strength & $B_\text{min}$ \\
        Particle distribution index & $p$ \\
        Size of corona & $H_\text{corona}$ \\
        Size of disk & $H_\text{disk}$ \\
        Size of halo & $H_\text{halo}$ \\
        \hline
    \end{tabular}
    \caption[List of free parameters for the simplified total field model]{List of free parameters for the simplified total field model.}
    \label{tab:parameters_dhc}
\end{table}

\begin{table}
\begin{tabular}{lcccc}
\hline
   & $H$/kpc & $L_c$/kpc & $z_{B}$/kpc & $z_{0}$/kpc \\
\hline
Disk & 0.2 & $5\times10^{-3}$ & $\infty$ & 5 \\
Halo & 10 & 10 & $5/3$ & 5 \\
Corona & 250 & $50\times10^{-3}$ & $5/3$ & 5 \\
        \hline
\end{tabular}
    \caption{Parameters of the simplified model.}
    \label{tab:DHC_setup}
\end{table}

To find a reasonable fit of the remaining parameters, we use a $\chi^2$
test to fit the latitude profile of the radio intensity at $\nu=408$\,Mhz.
We use the $N=14$  data points presented in Ref.~\cite{DiBernardo:2012zu}
together with the (asymmetric) errors for the radio intensity outside
the Galactic disk. We do not add the mean
polarization as an additional data point to the $\chi^2$ test: Treating
it as a single datum with a rather large variance, its weight in the fit
would be small. Instead, we calculate the mean polarization towards the
Galactic North pole for each fitted case, and consider then only cases below
a maximal polarization.

\subsection{Fitting  $p$, $z_{0}$ and $B_0$}

For our first $\chi^2$ test, we keep both the minimum electron number density
$n_\text{min}=0.1$ and the minimum magnetic field strength $B_\text{min}=0.5\mu$G
fixed. This leaves us with three free parameters; $p$, $z_{0}$ and $B_0$.
We choose seven equally spaced values for $p\in[2.4, 3]$, and five equally
spaced values for $z_{t,n}\in [2,6]$\,kpc and $B_0\in [3,7]\,\mu$G. 
At each latitude we average over $10^3$~field realizations.
The five best results of the test are summarized in
Table~\ref{tab:chi_square_test1}. The last column of the table also includes
the mean degree of polarization for the given parameters, calculated by the
line-of-sight integral towards the Galactic north pole.

\begin{table}[!htb]
    \centering
    \begin{tabular}{ccccccc}
        \hline
        $p$ & $z_0$ [kpc] & $B_0$ [$\mu$G] & $n_\text{min}$ & $B_\text{min}$ [$\mu$G] & $\chi^2$ & $\langle \Pi \rangle$ \\
        \hline
        2.7 & 2 & 6 & 0.1 & 0.5 & 0.22 & 0.15 \\
        2.7 & 4 & 5 & 0.1 & 0.5 & 0.41 & 0.17 \\
        2.7 & 3 & 5 & 0.1 & 0.5 & 0.47 & 0.15 \\
        2.7 & 6 & 4 & 0.1 & 0.5 & 0.63 & 0.17   \\
        2.7 & 3 & 6 & 0.1 & 0.5 & 0.67 & 0.17   \\
        \hline
    \end{tabular}
    \caption{Reduced $\chi^2$ values and mean degree of polarization for the five best set of parameters varying $p$, $z_0$ and $B_0$; the parameters $n_\text{min}$ and $B_\text{min}$ are fixed.}
    \label{tab:chi_square_test1}
\end{table}

\begin{figure}[!htb]
    \centering
    \includegraphics[width=\columnwidth]{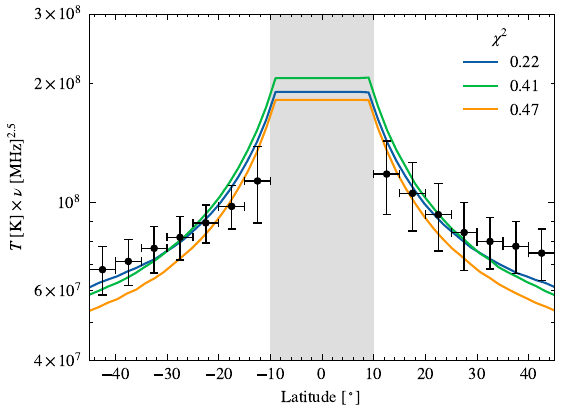}
    \caption{Latitude profiles of synchrotron emission for the simplified total field model for the three best results from Table~\ref{tab:chi_square_test1}.}
    \label{fig:intensity_latitude_dhc_test1}
\end{figure}

\begin{figure}
  \includegraphics[width=\columnwidth]{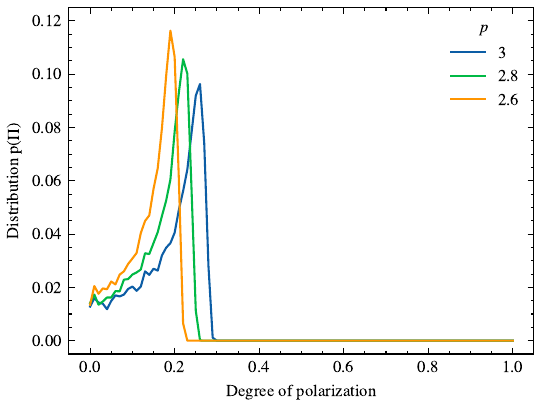}
\caption{Distribution of $p$ values for different slopes $p$ of the
  electron energy distribution, for $B_0=4\mu$G and $z_n=5$\,kpc.}
\label{fig:pol_dist}
\end{figure}

Values of $p$ close to $2.7$ are clearly favored by this test, while
the parameters $z_{0}$ and $B_0$ may vary more. The fact that $p$ is
well determined is partly a consequence of fixing
the normalization of the CR electron density. For the three best fits,
Fig.~\ref{fig:intensity_latitude_dhc_test1} shows the
total synchrotron radiation temperature $T=I_\nu c^2/(2\nu^2k_B)$
as function of the Galactic latitude $l$ together with the radio data.
%
%
The best fits for these model give mean polarization degrees
around $0.16$, which is still quite high. We check therefore how the
distribution of these values looks like.
In Fig.~\ref{fig:pol_dist}, we show the distribution of $\Pi$ values for
different realizations of the random field and three value of the slope
$p$ of the electron energy distribution. In all the cases shown in
Fig.~\ref{fig:pol_dist}, we fix $B_0=4\mu$G
and $z_0=5$\,kpc. The three distributions have a sharp
cutoff at high polarizations and a tail extending to $\Pi=0$.
Decreasing $p$,  the maxima and the cutoffs are shifted as expected
to smaller $\Pi$ values. For $p=2.8$, the probability
to have a random realization with polarization lower than $\Pi\leq 0.05$
is 9\%. Similarly, for the three best cases
from Table~\ref{tab:chi_square_test1} there are approximately 20\% of cases
with polarization less than $\Pi=0.1$ and 10\% of cases less than $\Pi=0.05$.
Thus we conclude that the extremely low polarization observed in the northern
C-BASS survey is not typical for our set-up, but happens in a considerable
fraction of field realizations.

\begin{figure*}
  \includegraphics[width=\columnwidth,angle=0]{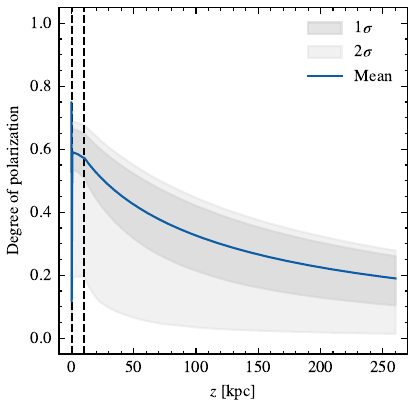}
  \includegraphics[width=\columnwidth,angle=0]{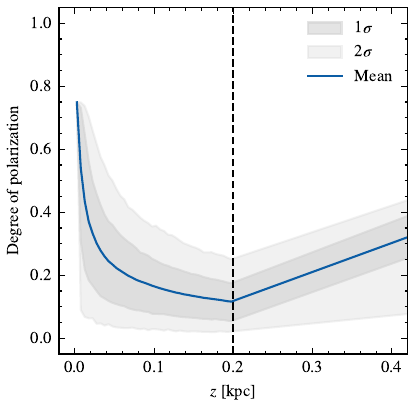}
\caption{Linear polarization degree $P$ as function of the L.o.S.\ distance
$s$.}
\label{fig:pol}
\end{figure*}

To understand this behavior better,  we show in
Fig.~\ref{fig:pol} the average of the linear polarization fraction $\Pi$
together with its 1 and $2\,\sigma$ bands as function of the line-of-sight
distance $s$. The left panel illustrates how the polarization degree is
reduced as $1/\sqrt{x}$, as the length $s=xL_c$ of the line-of-sight
integration increases in the turbulent extended halo field. The same effect
is visible for the turbulent disk field in the close-up shown in the right
panel.
Note also that the $2\,\sigma$ band is very asymmetrical: Depending on the
relative orientation of the polarization in the disk and the halo,
the two contributions can either cancel or sum up. In the first
case, the total polarization can become close to zero.
Thus very small values of $\Pi$ require a partial cancellation of the
disk and  halo contribution to the polarized emission.

\subsection{Fitting  $p$,  $n_\text{min}$ and $B_\text{min}$ }

In the second $\chi^2$ test, we want to determine the minimal contribution
to the synchrotron intensity required from the extended halo.
We therefore consider $n_\text{min}$ and $B_\text{min}$  together with $p$
as free fit parameters. Decreasing either   $n_\text{min}$ or $B_\text{min}$
increases the polarization degree and decreases the intensity, while 
on the other hand a smaller value of $p$ corresponds to a higher intensity
and lower polarization degree. We will therefore fix the disk field strength
$B_0 = 6\mu$G and the electron density scale height $z_0 = 2$\,kpc, and
vary $n_\text{min}$ and $B_\text{min}$ as well as $p$.

\begin{figure}
\includegraphics[width=\columnwidth,angle=0]{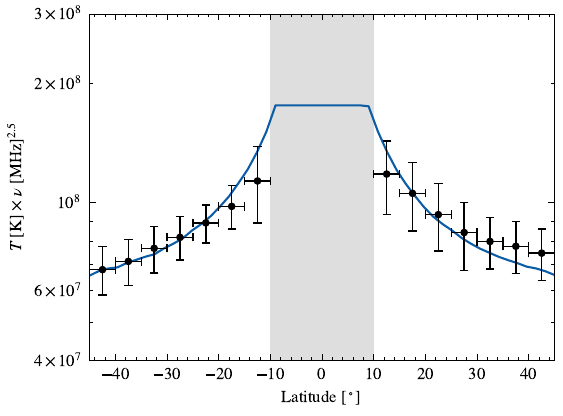}
\caption{Total synchrotron radiation temperature $T$ as function of
  Galactic latitude $l$.}
\label{fig:lat}
\end{figure}

Choosing six equally spaced values for $p\in[2.65, 2.7]$ and
$n_\text{min}\in [0.05,0.1]$, and three equally spaced values for
$B_\text{min}\in [0.3,0.5]\mu$G, gives the results of the $\chi^2$ test
presented in Table~\ref{tab:chi_square_test2_pol}. Here, we show the ten
latitude fits with a reduced chi-square value less than $0.5$; they
are sorted by the lowest mean polarization degree. All values obtained for
$n_\text{min}$, and $B_\text{min}$ are above the minimal values used in
the fit range. Thus the results of this fit point towards the need
of an extended halo field.
The synchrotron radiation temperature for the best-fit of the 18 models is
shown in Fig.~\ref{fig:lat} as function of the Galactic latitude.

\begin{table}[!htb]
  \begin{tabular}{ccccccc}
    \hline
    $p$ & $z_0$ [kpc] & $B_0$ [$\mu$G] & $n_\text{min}$ & $B_\text{min}$ [$\mu$G] & $\chi^2$ & $\langle \Pi \rangle$ \\
    \hline
    2.69 & 2 & 6 & 0.1 & 0.5 & 0.37 & 0.15 \\
    2.69 & 2 & 6 & 0.09 & 0.5 & 0.27 & 0.15 \\
    2.7 & 2 & 6 & 0.1 & 0.5 & 0.22 & 0.15 \\
    2.69 & 2 & 6 & 0.1 & 0.4 & 0.19 & 0.16 \\
    2.7 & 2 & 6 & 0.09 & 0.5 & 0.22 & 0.16 \\
    2.67 & 2 & 6 & 0.09 & 0.4 & 0.5 & 0.17 \\
    2.68 & 2 & 6 & 0.08 & 0.5 & 0.43 & 0.17 \\
    2.69 & 2 & 6 & 0.08 & 0.5 & 0.21 & 0.17 \\
    2.68 & 2 & 6 & 0.1 & 0.4 & 0.26 & 0.17 \\
    2.7 & 2 & 6 & 0.1 & 0.4 & 0.35 & 0.17 \\
    \hline
  \end{tabular}
  \caption{Reduced $\chi^2$ values and mean degree of polarization for varying $p$, $n_\text{min}$ and $B_\text{min}$ for the ten best sets of parameters with $\chi^2 < 0.5$ and lowest degree of polarization; the parameters $z_{0}$ and $B_0$ are fixed.}
        \label{tab:chi_square_test2_pol}
\end{table}

\section{Discussion}

We collect here few implications of our results and the model suggested.
\begin{itemize}
\item
In our fits, we have used  Eq.~(\ref{eq:n}) as functional form
for the vertical profile of the CR electron density together with
$j_0 z_{0}/(D_0 n_0) \simeq 2.0$ and the minimal value $n_{\min}=0.1$.
In a more physical way, the latter minimum could be implemented by choosing
$j_0 z_{0}/(D_0 n_0) \simeq 0.9$. Adjusting then also $z_0$ to  $z_0=0.5$\,kpc,
the profiles for these two choices agree for $|z|\to 0$. 
The combination $j_0/(D_0 n_0)$ of the remaining parameters differs only
by a factor of 2.3 between these two cases, indicating that 
a rather small change in, e.g., the CR luminosity can initiate Galactic
outflows which fill up the extended halo with CRs. For a recent study of such
a scenario, see Ref.~\cite{Zirakashivili:2023exz}.
\item 
Extended halos filled with CR electrons around other Milky Way-like galaxies
might lead to detectable radio halos. The intensity of these halos is however
low relative to the plane, preventing their detection with current
sensitivities~\cite{2018A&A...611A..72K}: As the intensity scales as
$I\propto L n B^2$, one expects a suppression of the halo intensity relative
to the peak by $(500/30)\times 0.1\times (0.5/6)^2 \simeq 0.01$
for a disk diameter of $L=30$\,kpc.
\item
Fitting observational data, the assumption that only the regular
magnetic field contributes to the polarized intensity and the
rotation measure, while the turbulent field cancels out completely,
does not hold necessarily for realistic conditions. For instance,
for $L_{\rm coh}=L_{\max}/5=10$\,pc in the disk, and a disk height of
$H= 150$\,pc, the turbulent components are suppressed only by
$1/\sqrt{15}\sim 0.25$. In reality, the suppression will be even
weaker, since the emissivity is not uniform.
Thus the regular and turbulent field should be fitted simultaneously to
observational data.
\item
Very small polarization fractions $\Pi$ require a partial cancellation of the
disk and halo contribution. In the case of regular fields, close to
perpendicular
field lines in and above the disk are however unlikely. In contrast,
if the turbulent component dominates in the disk (or at least
in the Local Bubble), such a partial cancellation happens naturally
in a subset of field realizations.
\item
An extended magnetic halo can influence the arrival directions and times
of ultra-high energy CRs. The rms deflection angle of an iron nucleus
in a turbulent field of length $d$ is given by~\cite{MiraldaEscude:1996kf}
\begin{align}
\label{eq:Ecr_sL}
\lefteqn{ \theta =  \frac{(2dL_c/9)^{1/2}}{R_L}}  
 \\ & \simeq 23^\circ  
      \left( \frac{10^{20}{\rm eV}}{E}\right)
      \left(\frac{B}{\mu{\rm G}}\right) 
 \left(\frac{d}{250\,{\rm kpc}}
 \frac{L_c}{50\,{\rm pc}}\right)^{1/2}, \nn
\end{align}
while the increased path-length compared to straight-line propagation leads
to time-delays 
\begin{align}
\label{eq:Ecr_sL2}
\Delta t & =  \frac{d\theta^2}{4} \simeq 1.3\text{Myr}  \times \\ &
 \left( \frac{10^{20}{\rm eV}}{E} \frac{B}{\mu{\rm G}} \right)^2
 \frac{d}{250\,{\rm kpc}} \frac{L_c}{50\,{\rm pc}} .
  \nn 
\end{align}
Thus the time delays are even at the highest energies for iron nuclei
smaller than the typical activity times of jets in AGN.
On the other hand, the halo field induces sizable additional deflections.
\item
The interactions of  CRs confined in the extended magnetic halo of the
Milky Way lead to a quasi-isotropic contribution to the photon and
neutrino fluxes~\cite{Taylor:2014hya}. We can estimate the ratio of this
secondary photon to the CR flux as
\be
\frac{I_\gamma(E)}{\vev{I_{\rm CR}(E)}}\simeq
H_{\rm cor} \vev{n_{\rm gas}} \sum_{ij} f_j Z_\gamma^{ij}(p)\sigma^{ij}_{\rm inel}(E),
\ee
where $f_j$ is the fraction of proton and helium nuclei in the interstellar
gas, $ Z_\gamma^{ij}(p)$ is the $Z$ factor for a power-law CR flux with
slope $p$, $\sigma^{ij}_{\rm inel}(E)$ the inelastic cross section for the
production for photons in interactions of CR nuclei of type $i$ on target
nuclei $j$, and $\vev{\cdot}$ denotes the average along the line-of-sight.
For a rough estimate, we set $\vev{n_{\rm gas}}=10^{-3}$cm$^{-3}$
as average gas density and $\vev{I_{\rm CR}(E,z)}=gI_{\rm CR}(E,0)$ with
$g\simeq n_{\min} = 0.1$. This results with $H_{\rm cor}=250$\,kpc and
$\sum_{ij} f_j Z_\gamma^{ij}(2.6)\sigma^{ij}_{\rm inel}(E)\simeq 7.3$\,mbarn
\cite{Kachelriess:2014mga}
in $I_\gamma(E)/I_{\rm CR}(E)\simeq  6\times 10^{-7}$.
Thus the secondary photons from CR interactions in the extended halo
could give a sizable contribution in the TeV range to the EGRB measured by
Fermi-LAT, for a potential signature in the data see
Ref.~\cite{Neronov:2018ibl}. However, in order to make
quantitative statements, more reliable estimates for
the gas and CR electron densities are needed.
A possible contribution to the diffuse
neutrino flux measured by IceCube would be stronger suppressed, first
because of the slope
$p=2.6$ and second because the knee in the Galactic CR flux is at an
$\simeq 10$ times lower energy than observed locally, as argued in
Ref.~\cite{Prevotat:2024ynu}.
\item  
CR Diffusion in the extended Milky Way halo leads to the formation of a
magnetic horizon, if
$\tau_{\rm esc}=H_{\rm cor}^2/(2D)>t_0$ with $t_0$ as the age of the Universe.
In a simplistic picture, the same condition determines the minimal rigidity
of extragalactic CRs which can enter the Milky Way.
With $\tau_{\rm esc}(\R)=t_0$, it follows $\R\gsim 10$\,PV. 
Thus a proton component at higher energies than $\simeq 10$\,PeV,
where the Galactic protons are already suppressed, might be extragalactic.
\item
Finally, we note that the different level of turbulence in the disk and
halo should lead to differences in the morphology of leptonic
gamma-ray sources: While in the disk such gamma-ray halos should be spherical,
halos of sources with a rather large vertical distance, $z\simeq 100$--200\,pc,
to the Galactic plane should be elongated along the regular magnetic field
lines.
%
\end{itemize}

\section{Conclusions}
%
We have argued that the low linear polarization degree in the northern
C-BASS survey together with coherent structures in the  polarization over
rather large angular scales is incompatible with a uniform ratio of the
regular and turbulent magnetic field strengths in the disk and halo of the
Milky Way. Since sufficiently fast CR escape requires the dominance of
regular fields in the halo, we have suggested the existence of two
additional regions dominated by turbulent fields: The disk---which may
be only locally dominated by turbulent fields---and  an extended halo field.
In the extended halo, the field strength and the density of CR electrons
should be sufficiently high that the field contributes significantly to
the observed synchrotron intensity. In the (local) disk, the ratio of
the coherence length and the vertical extension should be such that
only a partial cancellation of the disk contribution to the linear
polarization degree happens. 
The extended halo field may lead to sizable additional deflections
of UHECRs, while at lower energies interactions of confined Galactic
CRs may contribute to the observed diffuse flux of gamma-rays and
neutrinos.

\acknowledgments
This work was triggered by a talk of Paddy Leahy at the workshop
``Towards a comprehensive model of the Galactic magnetic field''  at
NORDITA, Stockholm. We would like to thank him and all participants
for inspiring discussions. We are grateful to Michael Unger for useful
comments on this draft.
MK acknowledges hospitality and financial support by NORDITA for
the scientific program ``Towards a comprehensive model of the Galactic
magnetic field'' and by the Bernoulli Center in Lausanne for
the scientific program ``Generation, evolution, and
observations of cosmological magnetic fields''.


\end{document}